\documentclass[aps,superscriptaddress,showpacs]{revtex4}
\usepackage{amssymb}
\begin{document}
\title{On thermodynamic and quantum fluctuations of cosmological
constant.}
\author{G.E. Volovik}
\email[]{volovik@boojum.hut.fi}
\affiliation{Low Temperature Laboratory, Helsinki University of Technology,
Box 2200, FIN-02015 HUT, Finland} 
\affiliation{L.D. Landau Institute for Theoretical
physics, 119334 Moscow, Russia.} 

\date{Version 5.00; 5 September 2004; \LaTeX-ed \today}
\begin{abstract}
We discuss from the condensed-matter point of view the recent idea that
the Poisson fluctuations of cosmological constant about zero could be a
source of the observed dark energy \cite{Padmanabhan,Ahmed}. We argue
that the thermodynamic fluctuations of $\Lambda$ are much bigger. Since
the amplitude of fluctuations is $\propto V^{-1/2}$ where $V$ is the
volume of the Universe, the present constraint on the cosmological
constant provides the lower limit for $V$, which is much bigger than the
volume within the cosmological horizon.  
\end{abstract}
\widetext
\maketitle


Recently the following approach to the solution of the cosmological
problem has been suggested: the average value of the cosmological constant
is zero, $\left<\Lambda\right>=0$, due to the ability of quantum micro
structure of spacetime to readjust itself and absorb bulk vacuum energy
densities; the observed dark energy is provided by residual quantum
fluctuations about zero which simulate a ``small'' cosmological constant
\cite{Padmanabhan,Ahmed}.

Here we discuss this scenario using the condensed-matter experience. In
the condensed-matter example of `quantum gravity',  the `trans-Planckian'
physics is well known, while gravity naturally emerges in the low-energy
corner together with gauge fields and chiral fermions (see
\cite{VolovikBook}). This working model of `quantum gravity' provides us
with some criteria for selection of theories: those theories of quantum
vacuum are favourable which are consistent with its condensed-matter
analog. Here we apply this test to the theory of the fluctuationg  
cosmological constant. We show that this idea is consistent
with the condensed-matter `oracle', but the amplitude of the fluctuations
is in contradiction. 

The first assumption made in papers \cite{Padmanabhan,Ahmed}, that
the average value of $\Lambda$ is zero, $\left<\Lambda\right>=0$, is
certainly consistent with the  condensed-matter `quantum gravity', where
the same cosmological constant problem arises and is resolved.  As in our
quantum vacuum, the naive estimation of the energy density of the
condensed matter in its ground state gives the `natural' characteristic
value for the induced cosmological constant
\begin{equation}
\Lambda_0\sim {E^4_{\rm Pl}\over \hbar^3 c^3}~.
\label{EstimatedLambda}
\end{equation}
Here $E_{\rm Pl}$ is the corresponding ultraviolet energy cut-off of
atomic scale, which plays the role of the Planck energy scale. However,
the microscopic (atomic) phyiscs is well known which allows us to make
exact calculations of the ground state energy. One obtains that for the
equilibrium ground state (vacuum) in the absence of excitations above the
vacuum (matter), the energy density of the system isolated from the
environment is exactly zero. In the language of gravity, this means that
$\left<\Lambda\right>=0$. The contribution $\Lambda_0$ from sub-Planckian
and Planckian modes of the quantum vacuum is exactly cancelled by the
microscopic trans-Planckian (atomic) degrees of freedom without any
fine-tuning.

This remarkable result follows from the general thermodynamic analysis
applied to the quantum vacuum, which utilizes the Gibbs-Duhem
relation (see e.g. Ref. \cite{Volovik3}).  The same analysis demomstrates
that in the presence of matter, $\Lambda$ is more or less on the order of
matter density or of other perturbations of the
vacuum \cite{VolovikBook,Volovik3}. 

If the perturbations of the vacuum are small, the cosmological
constant induced by these perturbations may become smaller than
thermodynamic or quantum fluctuations of the cosmological constant about
zero. Then we enter the regime, which has been considered in
Refs. \cite{Padmanabhan,Ahmed}. According to this suggestion the pesently
observed dark energy is nothing but the fluctuation of the cosmological
constant. To test this idea we extend our thermodynamic analysis
incorporating the thermodynamic and quantum fluctuations of $\Lambda$.

The thermodynamic fluctuations of cosmological constant $\Lambda$, and
thus of the vacuum pressure $P_{\rm vac}=-\Lambda$, correspond to the
thermodynamic fluctuations of pressure in condensed matter system, say
in a quantum liquid, when it is
close to its ground state (vacuum).  In quantum liquids,
assuming that the liquid is in equilibrium without
environment, the average pressure is zero
$\langle P\rangle=0$. This corresponds to 
\begin{equation}
 \langle
\Lambda\rangle=-\langle P_{\rm vac}\rangle=0 
\label{Zero}
\end{equation}
 in the equilibrium vacuum.
 The thermodynamic fluctuations of pressure are
given by the general thermodynamic relation:
\cite{StatPhys}
\begin{equation}
\langle P^2\rangle_{\rm Thermal}  
=-T\left({\partial P\over \partial V}\right)_S~,
\label{PressureFluctuations}
\end{equation}
where  $S$ is entropy, $T$ is temperature, and  $V$  is the volume of the
system.
In quantum liquids, the Eq.(\ref{PressureFluctuations}) is expressed
through  the mass density of the liquid  $\rho$ and the speed
of sound $s$:
\begin{equation}
\langle P^2\rangle_{\rm Thermal}  ={\rho s^2 T\over V}~,
\label{ThermodynamicFluctuations}
\end{equation}
Applying this to the vacuum, we use the fact that in quantum liquids, the
quantity
$\rho s^2$ is on the order of the characteristic energy density of the
liquid. That is why in the quantum vacuum, this would correspond to the
natural value of the vacuum energy density, i.e. to the natural value  of
the cosmological constant expressed through the Planck energy scale in
Eq.(\ref{EstimatedLambda}):
\begin{equation}
\rho s^2\equiv\Lambda_0\sim {E^4_{\rm Pl}\over \hbar^3 c^3}~,
\label{LambdaFluctuations}
\end{equation}
and we have the following estimation for the amplitude of thermal
fluctuations of the cosmological constant:
\begin{equation}
\langle \Lambda^2\rangle_{\rm Thermal}  = \langle P^2_{\rm
vac}\rangle_{\rm Thermal} = \Lambda_0{T\over V}~.
\label{LambdaThermalFluctuations}
\end{equation}
Thus the temperature of the Universe influences not only the matter
degrees of freedom (and thus the geometry of the Universe), but it
also determines the thermal fluctuations of cosmological constant. 
This is an important lesson from the condensed matter: thermal
fluctuations of the vacuum pressure are determined by vacuum degrees of
freedom, and not by much smaller matter degrees of freedom as one can
naively expect.

Note that in Eq.(\ref{LambdaThermalFluctuations}), $\Lambda_0$ is the
contribution to the vacuum energy density and to cosmological constant,
which comes from the Planck scale modes in the quantum vacuum. But both in
quantum liquids and in the quantum vacuum, the total contribution of all
the modes to the energy density is zero if the vacuum is in complete
equilibrium,
$\langle
\Lambda\rangle=0$. The tiny contribution to the vacuum energy density
comes from a small disbalance caused by perturbations of the vacuum
state, and here we discuss the disbalance produced by the
"hydrodynamic" fluctuations of the vacuum.

From Eq.(\ref{LambdaThermalFluctuations}) we can get the estimate for the
quantum fluctuations of the vacuum pressure and cosmological constant,
which are important when temperature becomes comparable to the distance
between the quantum levels in the finite volume, i.e. when $T\sim
\hbar c/V^{1/3}$. Substituting this $T$ into
Eq.(\ref{LambdaThermalFluctuations}) one obtains the amplitude of quantum
fluctuations of $\Lambda$:
\begin{equation}
\langle \Lambda^2\rangle_{\rm Quantum} =\langle P^2_{\rm
vac}\rangle_{\rm Quantum} =  \Lambda_0{\hbar c\over
V^{4/3}}~.
\label{LambdaQuantumFluctuations}
\end{equation}

Let us apply this first to the vacuum in the closed Einstein Universe with
the volume of the 3D spherical space $V=2\pi R^3$. This
gives the following estimation for the amplitude of quantum fluctuations
of
$\Lambda$:
\begin{equation}
 \sqrt{\langle \Lambda^2\rangle}_{\rm Quantum}  \sim
   {E^2_{\rm Pl}\over
\hbar c R^2}~.
\label{LambdaQuantumFluctuations2}
\end{equation}
It is comparable to the mean value of $\Lambda$ in Einstein Universe. For
example, in the cold Universe one has
$\langle\Lambda\rangle =  E^2_{\rm Pl}/8\pi \hbar c R^2$, while in the
hot Universe $\langle
\Lambda\rangle =  3E^2_{\rm Pl}/16\pi \hbar c R^2$. Moreover, in any
reasonable Universe the temperature is much bigger than the quantum
limit, $T\gg \hbar c/R$, and thus the thermodynamic fluctuations of
$\Lambda$ highly exceed the value required for the construction of the
closed solution. This means that  thermodynamic fluctuations would
actually destroy the Einstein solution. However, we must note that in the
thermodynamic derivation used it was assumed that the closed
Universe is in a thermal (but not in the dynamical) contact with the
`environment', i.e. the Universe is immersed in a thermal reservoir at a
fixed temperature, say, as a brane immersed in the higher-dimensional
world. The Einstein Universe under these conditions has been studied in
Ref.
\cite{Barcelo}. For the really closed system, i.e. without any contact
with the environment, such fluctuations are suppressed due to the energy
conservation.

Let us now turn to our Universe which is spatially flat and thus is
formally infinite. What volume is relevant for the thermodynamic and
quantum fluctuations?  If we suppose that
$V\sim R^3_h$, where
$R_h$ is on the order of the scale of the cosmological horizon, then for
the amplitude of the quantum  fluctuations of $\Lambda$ one obtains
just what has been suggested in Refs. \cite{Padmanabhan,Ahmed}:
\begin{equation}
\sqrt{\langle \Lambda^2\rangle}_{\rm Quantum} \sim {E^2_{\rm Pl}\over
\hbar c R^2_h}~.
\label{LambdaQuantumFluctuations2}
\end{equation}
For the same estimation of the vacuum energy density using different
approach see Ref. \cite{Bjorken}. 

Equation (\ref{LambdaQuantumFluctuations2}) gives the correct order of
magnitude for $\Lambda$ in the present Universe. However, again this is
negligibly small compared to the amplitude of thermal fluctuations of
$\Lambda$ at any resonable temperature of the Universe.
Assuming, for example, that the temperature of the Universe coincides
with the temperature
$T_{\rm CMB}$ of the cosmic microwave background radiation, one obtains
the following amplitude:
\begin{equation}
   \sqrt{\langle \Lambda^2\rangle}_{\rm CMB}    \sim
{E^2_{\rm Pl} T_{\rm CMB}^{1/2}\over
(\hbar c)^{3/2} R^{3/2}_h}\sim 10^{14} {E^2_{\rm Pl}\over
\hbar c R^2_h} \gg \sqrt{\langle \Lambda^2\rangle}_{\rm Quantum} ~.
\label{ThermalHorizon}
\end{equation}
Such a big value of $\Lambda$ is certainly in contradiction
with observations. This means that we cannot use the volume inside the
horizon as the proper volume for estimation of the quantum and thermal
fluctuations of
$\Lambda$. The reason is that we are interested not in the fluctuations 
of $\Lambda$ from one Hubble volume to another, but in the influence of
the $\Lambda$ term on the dynamics of the whole Universe.  Thus
instead of the Hubble volume $R^{3}_h$  one must use the real
volume of the Universe, which is not limited by the cosmological horizon.

From the condition
that thermal fluctuations of
$\Lambda$ at $T_{\rm CMB}$ must be smaller than the experimental upper
limit, $ \sqrt{\langle \Lambda^2\rangle}_{\rm CMB} <\Lambda_{\rm exp}
$, one can estimate the lower limit for the size of the Universe:
\begin{equation}
 V > {T_{\rm CMB} \Lambda_0\over\Lambda^2_{\rm exp}}\sim{T_{\rm CMB}
 \over \hbar H} R_h^3\sim 10^{28} R_h^3 ~,
\label{SizeUniverse}
\end{equation}
where $H\sim c/R_h$ is the Hubble constant.  Note that considering 
distances much bigger than the size of the horizon we do not sacrifice 
the usual notion of causality. The "hydrodynamic" fluctuations of the
vacuum are independent in different volumes of the space if these
volumes are separated by distance $|{\bf r}_1-{\bf r}_2|>\hbar c/T$:
\begin{equation}
\langle \delta P({\bf r}_1)\delta P({\bf r}_2)\rangle_{\rm Thermal} 
=\Lambda_0 T \delta({\bf r}_1-{\bf r}_2)~~, ~~|{\bf r}_1-{\bf r}_2|>{\hbar
c\over T}~.
\label{ThermodynamicFluctuationsCorrelator}
\end{equation}
Since $T\gg \hbar c/R_h$, the correlations decay at distances much
smaller than $R_h$, and thus we do not need correlations across the
horizon. 

The expanding Universe is certainly not in equilibrium. The noise caused
by deviations from the equilibrium adds to the estimated fluctuations and
may change the effective temperature. Moreover, $\Lambda$ in
Eq.(\ref{LambdaQuantumFluctuations2}) depends on time in expanding
Universe, and this dependence is inconsistent with observations.
Supernovae data show  that the dark energy density is either constant with
time  \cite{Riess}, or it still may depend on time, but rather weakly
\cite{Alam}. In the latter scenario, the dark energy density is constant
on average up to redshifts of 0.5 and growing to the past for larger
redshifts.  But it still remains significantly less than the matter
energy density.

Our thermodynamic analysis here cannot consider the dynamics of the
Universe. However, the fact that the dark
energy density is not very far from the density of dark matter
demonstrates that the Universe is not very far from equilibrium
\cite{Volovik3}, and thus our estimations are not highly distorted by
expansion. To incorporate fluctuations of the cosmological constant and
its time dependence in the real Universe, the Einstein equations must be
modified; otherwise both the fluctuations and time-dependence are
prohibited by Bianchi identities. But even in this case the
condensed-matter `quantum gravity' can be useful, since it provides us
with the general scheme of how the dissipation is introduced.
Following this scheme one can incorporates the extra terms the Einstein
equations which contain the time derivative of
$\Lambda$ \cite{Phenomenology}. The modification of the Einstein
equations was also suggested in
\cite{Ahmed}, but it represents the particular case of the two-parametric
modification suggested in
\cite{Phenomenology}. The two parameters
introduced in Ref. \cite{Phenomenology} and considered as 
phenomenological fitting parameters, can in principle fit the
real behavior of $\Lambda$ in expanding Universe.

And the last comment concerns the emergent gravity and
relativistic quantum field theory in the condensed-matter systems in
general. The usage of fermionic condensed matter as a primer of emergent
physics does not mean that we necessarily return to a view of spacetime as
an absolute background structure incorporating absolute simultaneity. The
emergence is based on the general topological properties of fermionic
vacua in momentum space  (see Chapter 8 of
\cite{VolovikBook}) which are not sensitive to the
background geometry: the background can be Galilean, Lorentzian, or
something else. The only requirement for
the application of the topology in momentum space is the invariance of
the vacuum under translations, so that the Green's function can be
represented in terms of the 4-momentum. If the Universe is
inhomogeneous, the 4-momentum and thus the momentum space topology are
well determined in the quasiclassical limit, i.e. for the wavelengths much
smaller than the characteristic size of the Universe. This is enough
for the construction of the universality classes of quantum vacua. For
one of the two important universality classes, the main physical laws
of Standard Model and gravity gradually emerge in the low energy corner
irrespective of the background. If the background is originally
non-Lorentzian, the information on this background is lost in the
low-energy corner where the effective Lorentzian space-time gradually
arises.  

It was recently suggested from the consideration of the highest energy
cosmic rays observed that Lorentzian invariance is probably more
fundamental than all the other physical laws, since it persists even well
beyond the Planck energy scale  $E_{\rm
Pl}$ \cite{Gagnon}. This is an encouraging 
fact for the emergent physics.  According to Bjorken
\cite{Bjorken2}, the emergence can explain the high precision of the
present physical laws only if there is an extremely small expansion
parameter. This small parameter can be the ratio of the energy scales:
$E_{\rm Pl}/E_{\rm Lorentz}$, where  $E_{\rm Lorentz}\gg E_{\rm Pl}$ is
the energy scale where the Lorentz invariance is violated.  In the
effective action for the gauge and gravity fields obtained by integration
over fermions, the integration region is limited by the Planck
energy cut-off. If 
$ E_{\rm Pl}\ll E_{\rm Lorentz}$, then integration is concentrated in the
fully relativistic region, where fermions are still very close to Fermi
points and thus obey the effective gauge invariance and general
covariance which follow from the topology of a Fermi point (see Chapter 33
of \cite{VolovikBook}). As a result the effective bosonic action also
becomes invariant, and the precision of the emergent physical laws is
determined by some power of $E_{\rm Pl}/E_{\rm Lorentz}$.

This is the reason why we assume that Lorentzian invariance is obeyed
at the microscopic level. The Lorentzian invariance implies that the
analog of the vacuum speed of sound $s$ in
Eq.(\ref{ThermodynamicFluctuations})  coincides with the speed of light
$c$. As a result the thermal fluctuations of $\Lambda$ are determined by
the "bare" vacuum energy
$\Lambda_0$ according to Eq. (\ref{LambdaFluctuations}).  

In conclusion, from the condensed-matter analog of `quantum gravity' it
follows that the main idea of Refs.
\cite{Padmanabhan,Ahmed}, that fluctuations
$\sqrt{\langle \Lambda^2\rangle}$ of
the cosmological cosnstant may dominate over its mean value ${\langle
\Lambda\rangle}$, is correct. However, it appears that the thermal
fluctuations of  $\Lambda$ at any reasonable temperature of the
Universe are much bigger than the `quantum' Poisson-type fluctuations
suggested in Refs. \cite{Padmanabhan,Ahmed}. If this analogy is correct,
the thermal fluctuations of $\Lambda$ pose the limit on the size of the
Universe, which must be by at least 9 orders of magnitude bigger than the
size of the cosmological horizon.

I thank J.D. Bjorken, P.B. Greene, T. Padmanabhan, L. Smolin, R.D.
Sorkin and A.A. Starobinsky for e-mail correspondence. This work is
supported in part by the Russian  Foundation for Fundamental Research
under grant $\#$02-02-16218; by the Russian Ministry of Education and
Science, through the Leading Scientific School grant $\#$2338.2003.2 and
the Research Program  ``Cosmion''; and by the
European Science Foundation Program COSLAB.


\end{document}